%% file: main.tex
\documentclass[10pt, conference, compsocconf]{IEEEtran}
\IEEEoverridecommandlockouts

\input{package}
\begin{document}
%
% paper title
% can use linebreaks \\ within to get better formatting as desired
\title{Pairwise Learning for Name Disambiguation in Large-Scale \\Heterogeneous Academic Networks}

\author{
\IEEEauthorblockN{Qingyun~Sun\IEEEauthorrefmark{1}\IEEEauthorrefmark{2},
Hao~Peng\IEEEauthorrefmark{1}\thanks{Qingyun Sun and Hao Peng contributed equally to this work.},
Jianxin~Li\IEEEauthorrefmark{1}\thanks{Jianxin Li is corresponding author.},
Senzhang~Wang\IEEEauthorrefmark{3}, 
Xiangyu~Dong\IEEEauthorrefmark{1},
Liangxuan~Zhao\IEEEauthorrefmark{1},
Philip~S.~Yu\IEEEauthorrefmark{4}
and
Lifang~He\IEEEauthorrefmark{5}}
\IEEEauthorblockA{\IEEEauthorrefmark{1}
Beijing Advanced Innovation Center for Big Data and Brain Computing,
Beihang University,
Beijing 100191, China}
\IEEEauthorblockA{\IEEEauthorrefmark{2}
Shenyuan Honors College,
Beihang University, 
Beijing 100191, China}
\IEEEauthorblockA{\IEEEauthorrefmark{3}
% College of Computer Science and Technology,
Nanjing University of Aeronautics and Astronautics, 
Nanjing 211106, China}
\IEEEauthorblockA{\IEEEauthorrefmark{4}
% Department of Computer Science,
University of Illinois at Chicago, 
Chicago 60607, USA
\IEEEauthorrefmark{5}
% Department of Computer Science and Engineering,
Lehigh University, 
Bethlehem, PA, USA}
\IEEEauthorblockA{
Email: \{sunqy, penghao, lijx\}@act.buaa.edu.cn, szwang@nuaa.edu.cn, 
\\
\{dongxiangyu, zhaolx\}@buaa.edu.cn, psyu@uic.edu, lih319@lehigh.edu}}
% make the title area
\maketitle

\begin{abstract}
Name disambiguation aims to identify unique authors with the same name. 
Existing name disambiguation methods always exploit author attributes to enhance disambiguation results. 
However, some discriminative author attributes (e.g., email and affiliation) may change because of graduation or job-hopping, which will result in the separation of the same author's papers in digital libraries.
Although these attributes may change, an author's co-authors and research topics do not change frequently with time, which means that papers within a period have similar text and relation information in the academic network. 
Inspired by this idea, we introduce Multi-view Attention-based Pairwise Recurrent Neural Network (\projtitle) to solve the name disambiguation problem. 
We divided papers into small blocks based on discriminative author attributes and blocks of the same author will be merged according to pairwise classification results of \projtitle.
\projtitle combines heterogeneous graph embedding learning and pairwise similarity learning into a framework. 
In addition to attribute and structure information, \projtitle also exploits semantic information by meta-path and generates node representation in an inductive way, which is scalable to large graphs. 
Furthermore, a semantic-level attention mechanism is adopted to fuse multiple meta-path based representations. 
A Pseudo-Siamese network consisting of two RNNs takes two paper sequences in publication time order as input and outputs their similarity. 
Results on two real-world datasets demonstrate that our framework has a significant and consistent improvement of performance on the name disambiguation task. 
It was also demonstrated that \projtitle can perform well with a small amount of training data and have better generalization ability across different research areas. 
\end{abstract}


% For peer review papers, you can put extra information on the cover
% page as needed:
% \ifCLASSOPTIONpeerreview
% \begin{center} \bfseries EDICS Category: 3-BBND \end{center}
% \fi
%
% For peerreview papers, this IEEEtran command inserts a page break and
% creates the second title. It will be ignored for other modes.
\IEEEpeerreviewmaketitle

\input{1_intro}
\input{2_relatedwork}
\input{3_method}
\input{4_experiment}
\input{5_conclusion}
\vspace{-0.4em}
\section*{Acknowledgment}
This work is supported by the the National Key R\&D Program of China (2018YFC0830804), NSFC No.61872022, NSF of Jiangsu Province BK20171420, NSF of Guangdong Province (2017A030313339) and CCF-Tencent Open Research Fund, and in part by NSF under grants III-1526499, III-1763325, III-1909323, and SaTC-1930941. 

% trigger a \newpage just before the given reference
% number - used to balance the columns on the last page
% adjust value as needed - may need to be readjusted if
% the document is modified later
%\IEEEtriggeratref{8}
% The "triggered" command can be changed if desired:
%\IEEEtriggercmd{\enlargethispage{-5in}}

% references section

% can use a bibliography generated by BibTeX as a .bbl file
% BibTeX documentation can be easily obtained at:
% http://www.ctan.org/tex-archive/biblio/bibtex/contrib/doc/
% The IEEEtran BibTeX style support page is at:
% http://www.michaelshell.org/tex/ieeetran/bibtex/
\bibliographystyle{IEEEtran}
% argument is your BibTeX string definitions and bibliography database(s)
\bibliography{IEEEabrv,ref}
%
% <OR> manually copy in the resultant .bbl file
% set second argument of \begin to the number of references
% (used to reserve space for the reference number labels box)
% \begin{thebibliography}{1}

% \bibitem{IEEEhowto:kopka}
% H.~Kopka and P.~W. Daly, \emph{A Guide to \LaTeX}, 3rd~ed.\hskip 1em plus
%   0.5em minus 0.4em\relax Harlow, England: Addison-Wesley, 1999.

% \end{thebibliography}

% that's all folks
\end{document}

%% file: package.tex
\usepackage{cite}
\let\labelindent\relax
\usepackage{enumitem}
\usepackage{amsmath,amssymb,amsfonts}
\usepackage{algorithmic}
\usepackage{graphicx}
\usepackage{textcomp}
\usepackage{xcolor}
\usepackage{multirow}
\usepackage{mathrsfs}
\usepackage{subfigure}
\usepackage{array}
\usepackage[ruled]{algorithm2e}
\definecolor{Gray}{gray}{0.95}
\usepackage{colortbl}
\usepackage{booktabs}
\definecolor{tabcolor}{rgb}{.0,.0,.1}
\usepackage{changes}
\usepackage{hyperref}
\newcommand{\projtitle}{MA-PairRNN\xspace}
\newcommand{\projtitleLSTM}{MA-PairRNN$_{\scriptscriptstyle \rm{LSTM}}$\xspace}

%% file: 1_intro.tex
\vspace{-0.6em}
\section{Introduction}
\label{sec:intro}
Namesake problem~\cite{han2004two} poses a huge challenge on many applications, e.g., information retrieval, bibliographic data analysis. 
When searching for academic publications by author name, the results may contain a long list of publications of multiple authors with the same name. 
Some digital libraries (e.g., DBLP\footnote{https://dblp.uni-trier.de/} and Google Scholar\footnote{https://scholar.google.com/}) list candidates after name disambiguation with corresponding homepage, email and affiliation to make it easier to obtain all publications of one particular author. 
The academic impacts of researchers are always measured by impacts of their publications in the research community. 
% Citation is always taken into consideration when it comes to funding and promotion. 
% In addition, users often search for researchers' names to keep up with their researches. 
Therefore, it is important to keep publication data in digital libraries accurate, consistent, and up to date. 

\begin{figure}[t]
% \vspace{-0.6em}
\centerline{\includegraphics[width=3.2 in]{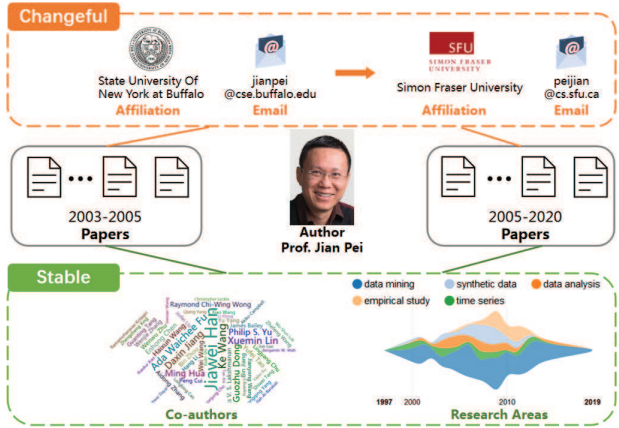}}
\caption{An example of the change of \textit{Jian Pei}'s discriminative attributes.}
\label{fig:background}
\vspace{-0.6em}
\end{figure}

Name disambiguation~\cite{fan2011graph,louppe2016ethnicity}, which aims to identify unique persons with the same name, has been studied for decades but remains largely unsolved. 
Most of the existing solutions utilize author attributes, including name, affiliation, email, homepage, etc., to generate paper representations or further validate disambiguation results. 
However, these discriminative attributes, especially email and affiliation, may change because of graduation or job-hopping. 
We take \textit{Jian Pei}, the well known leading researcher in data science, as an example to show the change of discriminative attributes in Fig.~\ref{fig:background}. 
\textit{Jian Pei}'s papers from 2003 to 2005 are associated with \textit{jianpei@cse.buffalo.edu} and \textit{State University of New York at Buffalo}. His papers from 2005 to 2020 are associated with \textit{jpei@cs.sfu.ca} and \textit{Simon Fraser University}. 
The change of discriminative attributes may lead to the paper separation problem~\cite{lee2005effective}, i.e., papers of an author are regarded as belonging to different authors, which commonly occurs in digital libraries. 
To address this issue, name disambiguation methods should perform well even when discriminative attributes change.

Even though discriminative attributes may have changed, researchers often have a fixed co-author set and a few specific research areas that do not change frequently over time, which can also be exploited to solve the name disambiguation problem. 
As shown in Fig.~\ref{fig:background}, even \textit{Jian Pei} has different affiliations and emails in two time periods, his close co-authors (e.g., \textit{Jiawei Han}, \textit{Ke Wang}) are fixed and his research areas (e.g., \textit{Data mining}, \textit{Time series}) are also consistent over time.

There are several challenges that should be overcome: 
\begin{figure}[t]
% \vspace{-0.6em}
\centerline{\includegraphics[width=3.2 in]{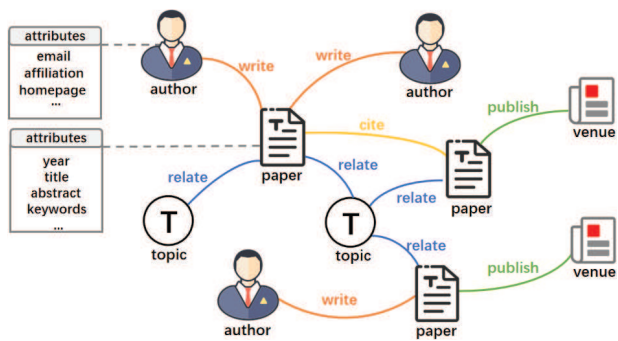}}
\caption{Academic network.}
\label{fig:network}
\vspace{-1 em}
\end{figure}
\begin{itemize}
\item[(1)]\textbf{Heterogeneity of academic network.} 
The academic network is a heterogeneous network that contains multiple entities (e.g., \textit{author}, \textit{paper}, \textit{venue}) and multiple relationships (e.g., \textit{writing}, \textit{publishing}) as shown in Fig.~\ref{fig:network}. 
It is challenging to preserve diverse structural and semantic information simultaneously.
\item[(2)]\textbf{Inductive capability.} 
Many real-world applications encounter a large number of new papers every day.
% , and the academic network evolves quickly. 
It is challenging for name disambiguation methods to have the inductive capability that can generate representations of new papers efficiently. 
\item[(3)]\textbf{Uncertain number of authors.} 
It is challenging to determine the number of authors with the same name. 
In existing clustering based name disambiguation methods~\cite{fan2011graph,louppe2016ethnicity,zhang2017name}, the number of authors (i.e., cluster size) is usually a pre-specified parameter.
\end{itemize}

Current works~\cite{zhang2018name,kim2019hybrid} did not efficiently handle the change of discriminative attributes and inductive paper embedding problem in the heterogeneous academic network simultaneously. 
In this work, we propose a novel \textbf{M}ulti-view \textbf{A}ttention-based \textbf{Pair} \textbf{R}ecurrent \textbf{N}eural \textbf{N}etwork framework, namely \textbf{\projtitle}, to solve name disambiguation problem.
The intuitive idea is that an author's papers during a period of time should have more similar representations since the co-authors and research interests of most authors are consistent despite attributes change. 
Inspired by this idea, we take name disambiguation as a pairwise paper set classification problem that does not require to estimate the number of authors with the same name. 
We divide papers into small blocks according to discriminative author attributes to reduce the search space of the name disambiguation algorithm. 
Then small blocks are merged based on pairwise classification result and each block after merging is the paper set of an author. 
We represent each paper block as a sequence in publication time order and solve the pairwise classification problem by comparing sequence similarity.
\projtitle combines multiple multi-view graph embedding layers, a semantic-level attention layer, and a Pseudo-Siamese recurrent neural network layer, to learn node embedding and node sequence pair similarity simultaneously. 
Specifically, multi-view graph embedding layer generates meta-path based embeddings of papers in the heterogeneous academic network. 
Then, semantic-level attention layer fuses these meta-path based embeddings into a vector. 
Finally, Pseudo-Siamese recurrent neural network layer learns the similarity of a node sequence pair. 
We elaborate on the three components as follows: 

\textbf{Multi-view graph embedding layer.} 
Multi-view graph embedding layer incorporates meta-paths to capture rich semantic information in the heterogeneous network. 
The heterogeneous network is converted into multiple relation view according to meta-paths. 
For each view, we learn $K$ aggregator functions to incorporate the $K$-hop neighborhood of each node. 
In this way, node embeddings are generated by enhancing node feature with semantics.

\textbf{Semantic attention layer.} 
% Different meta-path contains complex semantic information and plays different roles in downstream tasks. 
Semantic attention layer captures the importance of meta-paths by an attention mechanism and fuse semantic information for specific tasks. 
% It is important to select the most meaningful meta-paths and fuse semantic information for specific tasks. 
% We adopt an attention mechanism to capture the importance of meta-paths. 

\textbf{Pseudo-Siamese recurrent neural network layer.} 
Pseudo-Siamese recurrent neural network composes of two recurrent neural networks, which are used to learn inherent relations of paper sequences. 
It takes two sequences of paper embedding as input and outputs their similarity.

The main contributions are summarized as follows:
\begin{itemize}[leftmargin=*]
\item We propose a novel pairwise classification framework called \projtitle for name disambiguation task, which learns heterogeneous graph representation and paper set pairwise similarity simultaneously.
\item Under \projtitle, we propose an inductive graph embedding method that takes both heterogeneity and large scale of the academic network into account. 
A semantic-level attention mechanism is leveraged to put different emphases on each of the meta-paths. 
A Pseudo-Siamese recurrent neural network is adopted to learn inherent relations and measure the similarity of two paper sets.
\item We conduct extensive experiments on AMiner-AND and a large-scale real-world dataset collected from Semantic Scholar\footnote{https://www.semanticscholar.org/}. The results illustrate the best performance as well as good generalization ability of the proposed \projtitle compared to other methods. 
% \footnote{https://www.aminer.cn/na-data}
\end{itemize}
The code of \projtitle is available at \texttt{\url{https://github.com/RingBDStack/MA-PairRNN}}. 

%% file: 2_relatedwork.tex
\vspace{-0.6em}
\section{RELATED WORK}
In this section, we will briefly review name disambiguation methods and graph embedding methods. 
% We describe details as follows.
\subsection{Name Disambiguation}
Name disambiguation methods can be divided into supervised~\cite{han2004two,bunescu2006using}, unsupervised~\cite{zhang2018name,cen2013author} and graph-based ones~\cite{fan2011graph,zhang2017name}. 
Graph-based works exploit graph topological features in the academic network to enhance the representation of papers. 
For instance, GHOST~\cite{fan2011graph} constructs document graph based on co-authorship.
\cite{zhang2017name} leverages only relational data in the form of anonymized
graphs to preserve author privacy. 
Pairwise classification methods are applied to estimate the probability of a pair of author mentions belonging to the same author and are essential in the name disambiguation task.
\cite{zhang2018name} first learns representation for every name mention in a pairwise or tripletwise way and refines the representation by a graph auto-encoder, but this method neglects linkage between paper and author and co-authorship. 
\cite{kim2019hybrid} addresses the pairwise classification problem by extracting both structure-aware features and global features without considering semantic features. 
In this paper, we focus on the paper set level pairwise classification problem and exploit attribute, structure, and semantic features to form better representation.

\subsection{Graph Embedding}
Graph embedding aims to represent a graph as a low dimensional vector while preserving graph structure and properties. 
Recently, Graph Neural Network (GNN)~\cite{wu2020comprehensive} has attracted rising attention due to effective representation ability. 
% captures semantic and structure information through message passing between nodes. 
While most GNN works~\cite{wu2020comprehensive,peng2018large,peng2019hierarchical} focus on transductive setting, there have been some recent works adopting an inductive learning setting. 
DeepGL~\cite{rossi2017deep} aggregates a set of base graph features by relational functions that can generalize across networks. 
GraphSage~\cite{hamilton2017inductive} samples a fixed number of neighbors and generate node embeddings by aggregating their features. 
Both DeepGL and GraphSage are designed for homogeneous graphs. 
LAN~\cite{wang2019logic} aggregates neighbors with both rule-based and network-based attention weights for knowledge graphs. 

% Most of these methods neglect the heterogeneous semantics in data and cannot scale to large graphs.
Heterogeneous information networks~\cite{wang2014mmrate,zhang2018improving,gao2020adversarialnas,cao2020multi} have been studied in recent years. 
Meta-path is designed to preserve diverse semantic information of node type and edge type~\cite{sun2011pathsim,peng2019fine,he2019hetespaceywalk}. 
GTN~\cite{yun2019graph} converts heterogeneous graph to new graph structures which involve identifying task-specific meta-paths and multi-hop connections. 
HAN~\cite{wang2019heterogeneous} includes both node-level and semantic-level attention to take the importance of nodes and meta-paths into consideration simultaneously. 

In this paper, we propose an inductive graph embedding method utilizing rich heterogeneous information. 
% \subsection{Pseudo-Siamese Network}
% Siamese networks~\cite{bromley1994signature,baldi1993neural} are a family of networks that consist of two networks with shared parameters and a small network to fuse the output of two networks. 
% Siamese networks integrate representation learning and similarity learning into a framework and have achieved great success in visual recognition tasks~\cite{koch2015siamese,bertinetto2016fully}. 
% As a variation of the Siamese network, the Pseudo-Siamese network consists of two networks, which can be of the same or different types of neural networks (such as LSTM and CNN) and do not share parameters. 
% Recurrent neural network (RNN)~\cite{mikolov2010recurrent} has enjoyed rich success in many time series analyzing problems~\cite{gers2002applying,malhotra2015long}. 
% In this paper, we adapt the Pseudo-Siamese network with two RNNs to handle paper sequence pair similarity learning problems. 

%% file: 3_method.tex
\vspace{-0.3em}
\section{Proposed Method}
\subsection{Problem Definition}
In this section, we formally define Heterogeneous Academic Network and the problem of Name Disambiguation.
\newtheorem{definition}{Definition}
\begin{definition}[\textbf{Heterogeneous Academic Network}]
Heterogeneous Academic Network is defined as $\mathcal{G=\{V, E\}}$, where $\mathcal{V}$ and $\mathcal{E}$ denote the set of nodes and edges, respectively. 
A Heterogeneous Academic Network is associated a node type mapping function $f_v:\mathcal{V \rightarrow O}$ and an edge type mapping function $f_e:\mathcal{E \rightarrow R}$. 
$\mathcal{O}=\{P, A, T, V\}$ denotes node types set and $\mathcal{R}$ = \{\textit{A~writes~P}, \textit{P~cites~P}, \textit{P~is~related~to~T}, \textit{P~is~published~in~V}\} denotes edge types set, where $P, A, T, V$ denote the type of \textit{Paper}, \textit{Author}, \textit{Topic} and \textit{Venue}, respectively.
\end{definition}
\begin{definition}[\textbf{Name Disambiguation}]
% Name disambiguation aims to find unique authors from all author mentions extracted from papers or records in digital library. 
Given a name $a$, $\mathcal{D}^a=\{d^a_{1}, d^a_{2}, \dots, d^a_{N}\}$ is a set of papers with name mention $a$. Every paper $d^a_{k}$ consists of some metadata including paper attributes (e.g. \textit{title}, \textit{year}, \textit{venue}, \textit{keywords}) and author attributes (e.g. \textit{name}, \textit{email}, \textit{affiliation}). The objective of name disambiguation is to partition all name mentions into a set of unique authors $\mathcal{C}^a=\{c^a_{1},c^a_{2},\dots,c^a_{n}\}$. 
\end{definition}
\subsection{Model Architecture}
\begin{figure*}[t]
\centerline{\includegraphics[width=6.9 in]{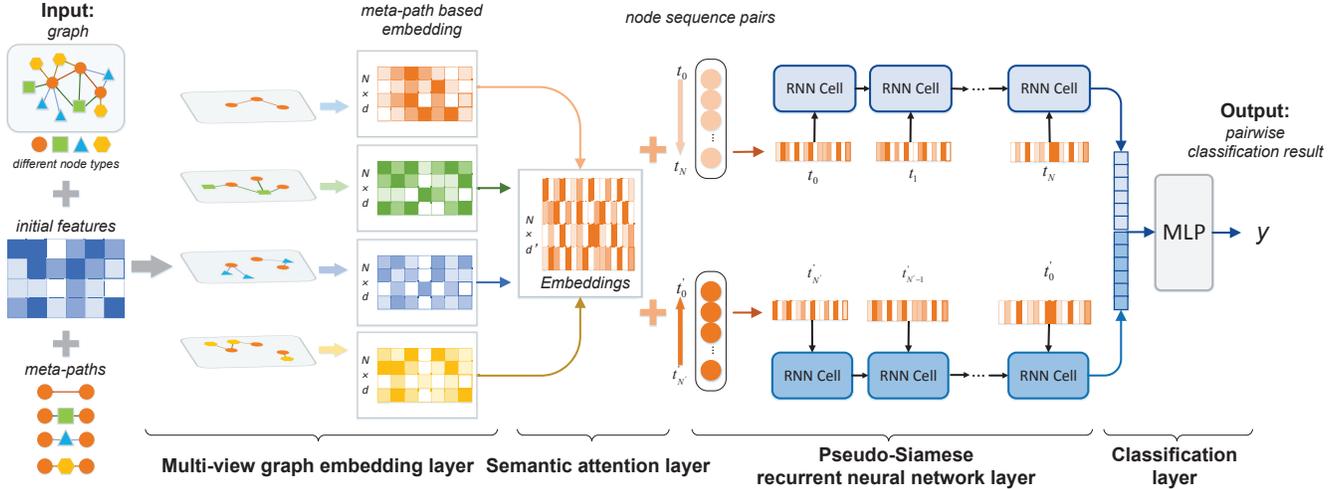}}
\caption{An overview of our overall network architecture.}
\label{fig:architecture}
\end{figure*}

In this section, we propose a novel framework named \textbf{\projtitle} for name disambiguation. 
As described above, the main intuition is that papers of the same author within a period should have similar representations in the academic network since the author's research and scholar relation is consistent. 
We divide the set of papers $\mathcal{D}^a$ into small blocks by discriminative author attributes in metadata. 
These small blocks will be merged based on pairwise classification results of \projtitle. 
First, the multi-view inductive graph embedding layer is designed to generate the paper representation of each meta-path. 
Then a semantic attention layer is designed to learn importance of meta-paths and fuse meta-path based representations.
Finally, papers in every block are arranged as a sequence denoted as $s\in\mathcal{S}$ according to their publication time. 
Two sequences of paper embedding are fed into a Pseudo-Siamese network with two RNNs for pairwise similarity learning. 
The overall architecture of \projtitle is shown in Fig.~\ref{fig:architecture}
\subsection{Multi-View Graph Embedding Layer}
Multi-view graph embedding layer generates node representations inductively by learning a function to aggregate attribute and topology information from local neighborhoods. 
To exploit rich semantic information in the heterogeneous academic network, we proposed the concept of meta-path based view. 
Given a heterogeneous academic network $\mathcal{G=\{V,E\}}$ and a meta-path $p$, a meta-path based view $\mathcal{G}_p$ is derived from a type of proximity or relationship between nodes characterized by a meta-path. 
It can capture different aspects of structure information through meta-paths and is potential to add new nodes dynamically.

% For each meta-path based view, similar to GraphSAGE \cite{hamilton2017inductive}, node representations are generated by stacking $K$ graph embedding layers, and in each layer an aggregator function aggregates features of meta-path based neighbors and propagates information across different layers. 
For each meta-path based view, similar to GraphSage~\cite{hamilton2017inductive}, node representations are generated by aggregating features of meta-path based neighbors and propagating information across $K$ layers. 
Node $v$'s representation based on meta-path $p$ is generated as below. 
First, in the $k$-th layer, each node aggregates its own representation and representations of its 1-hop neighborhood $\mathcal{N}_i$ generated by ($k$-1)-th layer into a single vector $\mathbf{z}^{(k)}_{p}(\mathcal{N}_i)$ as \eqref{eq:aggregate}:
\begin{equation}
\label{eq:aggregate}
 \mathbf{z}^{(k)}_{p}(\mathcal{N}_i)=mean(\{\mathbf{z}^{(k-1)}_{p}(v_j), \forall  v_j \in v_i\cup\mathcal{N}_i\}),
\end{equation}
where $\mathbf{z}^{(k-1)}_{p}(v_j)$ denotes representation of $v_j$ in ($k$-1)-th layer. 
When $k$ = 0, $\mathbf{z}^{(0)}_{p}(v_j)$ is defined as original feature $x(v_j)$ of $v_j$. 
Then a weight matrix $\mathbf{W}^{(k)}_p$ and a bias vector $\mathbf{b}^{(k)}_p$ are used to transfer information between layers as \eqref{eq:sage}:
\begin{equation}
\label{eq:sage}
\mathbf{z}^{(k)}_{p}(v_i)=\sigma(\mathbf{W}^{(k)}_p \cdot \mathbf{z}^{(k-1)}_{p}(\mathcal{N}_i)+\mathbf{b}^{(k)}_p).
\end{equation}

To extend the algorithm to a mini-batch setting, we first sample the $l$-egonet of papers in the batch. 
The $l$-egonet of node $v$ is defined as the set of its $l$-hop neighbors and all edges between nodes in the set. 
For each batch, multi-view subgraphs are constructed based on the union of $l$-egonets of all paper nodes in this batch. 
Then we generate meta-path based representation of every node in these multi-view subgraphs.
For more convenient notation, we denote $v_i$'s final representation based meta-path $p$ after $K$ layers as $\mathbf{z}_{p}(v_i)\equiv\mathbf{z}^{(K)}_{p}(v_i)$, where $\mathbf{z}_{p}(v_i)\in\mathbb{R}^{{d}}$.

\subsection{Semantic Attention Layer}
For each paper, multiple meta-path based representations are obtained and they can collaborate with each other. 
Since we assume that the importance of meta-paths varies, an attention mechanism is adopted to capture their contribution and fuse meta-path based node representations.

% Following~\cite{zhou2019hahe}, 
We first introduce a meta-path preference vector $\mathbf{a}_{p}\in\mathbb{R}^{\left | \mathcal{P}\right |\ast {d}'}$ for each meta-path $p$ to guide the semantic attention mechanism. 
For meta-path based representation $\mathbf{z}^{(k)}_{p}$ and meta-path preference vector $\mathbf{a}_{p}$, the more similar they are, the greater weight will be assigned to $\mathbf{z}^{(k)}_{p}$. 
We use a non-linear function to transform the $d$-dimension meta-path based embedding into ${d}'$-dimension as \eqref{eq:att_trans}:
\begin{equation}
\vspace{-0.3em}
\label{eq:att_trans}
\mathbf{z}'_{p}(v_i)=\sigma(\mathbf{W}_{p}\cdot\mathbf{z}_{p}(v_i)+\mathbf{b}_{p}).
\end{equation}
where $\mathbf{W}_{p}\in\mathbb{R}^{\left | \mathcal{P}\right |\ast {d}'}$ is the  weight parameter and $\mathbf{b}_{p}\in\mathbb{R}^{{d}'}$ is the bias parameter of transformation. 
$\mathbf{z}'_{p}(v_i)\in\mathbb{R}^{{d}'}$ is the node representation of $v_i$ based meta-path $p$ after transformation. 
The similarity of transformed representation vector and preference vector $\omega_{p}(v_i)$ is calculated as \eqref{eq:att_sim}:
\begin{equation}
\vspace{-0.1em}
\label{eq:att_sim}
\omega_{p}(v_i)=\frac{\mathbf{a}^{T}_{p}\cdot\mathbf{z}'_{p}(v_i)}{\Vert\mathbf{a}_{p} \Vert \cdot \Vert\mathbf{z}'_{p}(v_i)\Vert},
\vspace{-0.3em}
\end{equation}
where $\Vert\cdot\Vert$ is the \textit{L2} normalization of vectors. 
The weight of meta-path $p$ for node $v_i$ is defined using a softmax unit as follows:
\begin{equation}
\vspace{-0.3em}
\label{eq:att_weight}
\omega'_{p}(v_i)=\frac{exp(\omega_{p}(v_i))}{\sum_{{{p}'}\in\mathcal{P}}{exp(\omega_{{p}'}(v_i))}}.
\end{equation}

Final representation of node $v_i$ is generated by fusing all meta-path based representations in the weighted sum form:
\begin{equation}
\label{eq:att_sum}
\mathbf{z}(v_i)=\sum_{{p}'\in\mathcal{P}}{\omega'_{{p}'}(v_i)\ast \mathbf{z}_{{p}'}(v_i))}.
\end{equation}

\subsection{Pseudo-Siamese Recurrent Neural Network Layer}
We designed a Pseudo-Siamese recurrent neural network layer to capture inherent relations of papers and measure similarity of two paper sets. 
Pseudo-Siamese recurrent neural network layer is a Pseudo-Siamese network consisting of two RNNs with different parameters to generate representations of two node sequences. 
% LSTM is a special kind of RNN with memory cells and forget units, capable of learning long-term dependencies. 
% We combine the commonly used equations with our settings. 
% The update equations of both LSTMs in the Pseudo-Siamese network are as follows:
% \begin{equation}
% \label{eq:LSTM}
% \begin{aligned}
% \mathbf{i}_{t} &= \sigma\left(\mathbf{W_{i i}} \mathbf{z}_{t}+\mathbf{b_{i i}}+\mathbf{W_{h i}} \mathbf{h}_{(t-1)}+\mathbf{b_{h i}}\right), \\
% \mathbf{f}_{t} &= \sigma\left(\mathbf{W_{i f}} \mathbf{z}_{t}+\mathbf{b_{i f}+W_{h f}} \mathbf{h}_{(t-1)}+\mathbf{b_{h f}}\right), \\
% \mathbf{g}_{t} &= \tanh \left(\mathbf{W_{i g}} \mathbf{z}_{t}+\mathbf{b_{i g}+W_{h g}} \mathbf{h}_{(t-1)}+\mathbf{b_{h g}}\right), \\
% \mathbf{o}_{t} &= \sigma\left(\mathbf{W_{i o}} \mathbf{z}_{t}+\mathbf{b_{i o}+W_{h o}} \mathbf{}_{(t-1)}+\mathbf{b_{h o}}\right), \\
% \mathbf{c}_{t} &= \mathbf{f}_{t} * \mathbf{c}_{(t-1)}+\mathbf{i}_{t} * \mathbf{g}_{t}, \\
% \mathbf{h}_{t} &= \mathbf{o}_{t} * \tanh \left(\mathbf{c}_{t}\right),
% \end{aligned}
% \end{equation}
% where $\mathbf{z}_t$ represents the representation vector of $t$-th node in sequence and it can be calculated by \eqref{eq:att_sum}. 
% And $\mathbf{h}_{t-1}$ is the $(t-1)$-th hidden output vector. 
% The cell state $\mathbf{c}_t$ is updated by previous state $\mathbf{c}_{t-1}$ controlled by forget gate $\mathbf{f}_t$ and new input value $\mathbf{g}_t$ controlled by input gate $\mathbf{i}_t$. 
% Final output $\mathbf{h}_t$ is based on current cell state $\mathbf{c}_t$ and controlled by output gate $\mathbf{o}_t$. 
Specifically, we feed two sequence of paper embeddings into two RNNs respectively. 
The learned paper embedding of the paper is taken as the input of RNN units. 
The output of each RNN unit can be formalized as: 
\begin{equation}
\label{eq:RNN}
    \mathbf{h}_{t} = \rm{RNN}(\mathbf{z}_t, \theta_{t}),
\end{equation}
where $\theta_{t}$ means parameters of RNN unit. 
Here we apply the popular LSTM to capture inherent relations of paper sequences and learn their similarity. 
% In preliminary experiments, we also tried with other RNNs like BiLSTM~\cite{huang2015bidirectional} and GRU~\cite{cho2014learning}, but found LSTM more stable. 
Note that the paper sequence published earlier is in published time order and the other sequence is in reverse. 
This setting is based on the assumption that an author's research topics and co-authors are stable during the period of attribute changing. 
All outputs of RNN units are aggregated by a $GlobalPool$ function to generate the representation of paper sequence as follows:
\begin{equation}
\vspace{-0.3em}
\label{eq:seq_mean}
\mathbf{h} = GlobalPool(\{\mathbf{h}_{t},t=1,2,\cdots,| s|\}),
% \begin{aligned}
% \mathbf{h}^{(1)} &= READOUT(\{\mathbf{h}^{(1)}_{t},t=1,2,\cdots,| s^{(1)}|\}),\\
% \mathbf{h}^{(2)} &= READOUT(\{\mathbf{h}^{(2)}_{t},t=1,2,\cdots,| s^{(2)}|\}),
% \end{aligned}
\vspace{-0.1em}
\end{equation}
where $|\cdot|$ denotes the length of sequence. 
We apply a simple averaging strategy as the ${GlobalPool}$ function here. 
The final representations of two paper sequences $\mathbf{h}^{(1)}$ and $\mathbf{h}^{(1)}$ are concatenated and then fed into a multiple fully connected neural network:
\begin{equation}
\vspace{-0.3em}
\label{eq:mlp}
\mathbf{\hat{y}}_s = \sigma(\rm{MLP}([\mathbf{h}^{(1)},\mathbf{h}^{(2)}])),
\end{equation}
where $\sigma(\cdot)$denotes the softmax function and $[\cdot,\cdot]$ represents the concatenation operation.

Since our task is classification, the loss function $\mathcal{L}_{classify}$ can be defined as the Cross-Entropy over all labeled node sequence pairs between the ground-truth and the predict results. 
% \begin{equation}
% \vspace{-0.3em}
% \label{eq:cross}
% \begin{aligned}
% \mathcal{L}_{classify} &= -\sum_{s\in\mathcal{S}}\mathbf{y}_s\ast\log(\mathbf{\hat{y}}_s),
% \end{aligned}
% \end{equation}
% where $\mathbf{y}_s$ is the true label of node sequence pair $s$.
The proposed framework can be trained on a set of example pairs. 
For each pair of paper sequences, a cosine score function is applied to measure the similarity of the two paper sequence representations as \eqref{eq:cosine}. 
\begin{equation}
\label{eq:cosine}
\mathcal{L}_{sim} = sim(\mathbf{h}^{(1)},\mathbf{h}^{(2)}) = \frac{\mathbf{h}^{(1)}\cdot\mathbf{h}^{(2)}}{\left \|\mathbf{h}^{(1)} \right \|\cdot\left \|\mathbf{h}^{(2)} \right \|}.
\end{equation}
The pairwise similarity loss function encourages node sequences of the same author to have similar representations, and enforces that of different authors to be highly distinct.

The model is then trained to minimize the sum of classification loss as follows:
\begin{equation}
\label{eq:loss}
\mathcal{L}=\mathcal{L}_{classify}+\eta *\mathcal{L}_{sim},
\end{equation}
where $\eta$ denotes the coefficient of pair similarity loss.
The overall process of \projtitle is shown in Algorithm~\ref{alg:1}.
\begin{algorithm}[t]
\caption{The overall process of \projtitle}\label{alg:1}
\LinesNumbered
\KwIn{Paper set $\mathcal{D}$, heterogeneous graph $\mathcal{G=\{V,E\}}$, node features $\{\mathbf{x}(v),\forall v \in \mathcal{V}\}$, meta-path set $\mathcal{P}=\{p_1,p_2,\cdots,p_M\}$, number of multi-view graph embedding layer $K$}
\KwOut{meta-path based node representation $\{\mathbf{z}_{p_1},\mathbf{z}_{p_1},\cdots,\mathbf{z}_{p_1}\}$}
Separate paper set $\mathcal{D}$ into small blocks according discriminative author attributes;

Arrange papers in every block as sequence $s\in\mathcal{S}$;

Construct meta-path based view $\{\mathcal{G}_{p_1},\mathcal{G}_{p_2},\cdots,\mathcal{G}_{p_M}\}$;

$\mathbf{z}^{(0)}_{p}(v_i) = \mathbf{x}(v_i) , \forall v_i \in \mathcal{V};$

\While{not converge}{
    \For{$v_i \in \mathcal{V}$}{
        \For{$p \in \mathcal{P}$}{
            \For{$k=1,2,\cdots,K$}{
                Aggregate meta-path based neighbor information in previous layer by \eqref{eq:aggregate};
                
                Calculate the representation of current layer by \eqref{eq:sage};
            }
        }
        Calculate the attention weight of each meta-path by \eqref{eq:att_trans}, \eqref{eq:att_sim}, \eqref{eq:att_weight};
        
        Fuse the semantic representation of each meta-path based view by \eqref{eq:att_sum};
    }
    \For{$s\in\mathcal{S}$}{
        Calculate the representation of sequence pair by \eqref{eq:RNN} and \eqref{eq:seq_mean};
        
        Classify the sequence pair by \eqref{eq:mlp};
    }
    Calculate Loss by \eqref{eq:cosine} and \eqref{eq:loss}.
    
    % Back propagate and update parameters.
}
\end{algorithm}

%% file: 4_experiment.tex
\vspace{-0.3em}
\section{EXPERIMENTS}
\subsection{Dataset}
\begin{figure}[t]
\vspace{-0.6em}
\centerline{\includegraphics[width=3.2in]{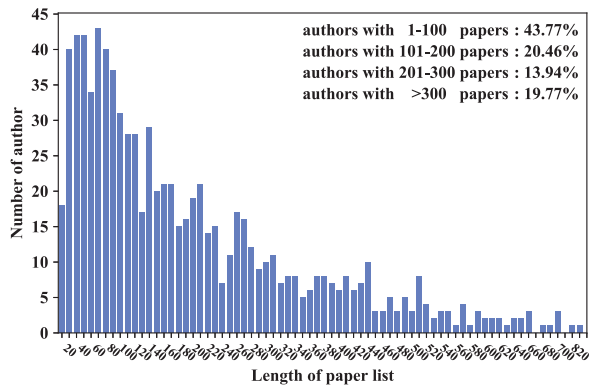}}
\caption{Length Statistics of Paper sets.}
\label{fig:list_len}
\vspace{-0.6em}
\end{figure}

For our experiments we used two datasets: Aminer-AND and Semantic Scholar.
\begin{itemize}[leftmargin=*]
    \item \textbf{Aminer-AND}~\cite{zhang2018name}: This dataset contains 70,285 records of 12,798 unique authors with 100 ambiguous name references.
    \item \textbf{Semantic Scholar}: We construct a new real-world academic dataset from a digital library called Semantic Scholar.
    There are 154,822 records of 857 unique authors with 226 highly ambiguous name in medicine area and reference papers of these records. 
    Detailed description is shown in Table~\ref{table:dataset}.
    The statistics of these authors' papers are shown in Fig.~\ref{fig:list_len}. 
    % We can see that it follows the long-tailed distribution and more than 64\% authors have papers within 200.
\end{itemize}
\begin{table}[t]
\caption{Statistics of Semantic Scholar}
\label{table:dataset}
\vspace{-0.6em}
\begin{center}
\begin{tabular}{c|lr|lr}
\hline
\textbf{Dataset}              & \textbf{Node Types} & \textbf{\#Nodes}   & \textbf{Relation Types} & \textbf{\#Edges}   \\ \hline
\multirow{4}{*}{\begin{tabular}[c]{@{}c@{}}Semantic\\Scholar\end{tabular}} & \textit{author}     & 1,891,542 & \textit{author-paper} & 4,607,109 \\
                     & \textit{paper}      & 698,219   & \textit{paper-term}   & 7,713,923 \\
                     & \textit{topic}      & 135,596   & \textit{paper-venue}  & 5,21,601  \\
                     & \textit{venue}      & 26,160    & \textit{paper-paper}  & 929,429   \\ \hline
\end{tabular}
\end{center}
\vspace{-0.6em}
\end{table}
\subsection{Evaluation Metrics and Baselines}
We apply pairwise Precision, Recall and F1 score in Aminer-AND and apply averaged Accuracy, F1 score and AUC in Semantic Scholar to measure the performance of all methods. 
We compare with attribute based methods as well as attribute and structure based methods to demonstrate the effectiveness of our model. 
To verify the effectiveness of each component including meta-path based views, semantic-level attention and Pseudo-Siamese structure, we also test three variants of \projtitle.
\begin{itemize}[leftmargin=*]
\item \textbf{MLP}\cite{pal1992multilayer}: It's s multilayer perceptron that directly projecting input features into a low dimensional vector.
\item \textbf{Deepwalk}\cite{perozzi2014deepwalk}: Deepwalk captures contextual information of neighborhood via uniform random walks for node embedding in homogeneous network.
\item \textbf{GraphSage}\cite{hamilton2017inductive}: GraphSage samples node neighborhoods to generate node embeddings for unseen data in an inductive way and is designed for homogeneous network.
\item \textbf{Zhang et al.}\cite{zhang2017name}: This method learns paper embedding by sampling triplets from three graphs constructed by relations of authors and papers and cluster them by hierarchical agglomerative algorithm.
\item \textbf{GHOST}\cite{fan2011graph}: GHOST use affinity propagation algorithm for clustering on a co-authors graph where the node distance is measured based on the number of valid paths. 
\item \textbf{Louppe et al.}\cite{louppe2016ethnicity}: This method trains a pairwise distance function based on similarity features and use a semi-supervised HAC algorithm for clustering.
\item \textbf{Aminer}\cite{zhang2018name}: This method first learns supervised global embeddings and then refines the global embeddings for each candidate set based on the local contexts.
\item \textbf{Kim et al.}\cite{kim2019hybrid}: It is a hybrid pairwise classification method which generates paper representation by extracting both structure-aware features and global features.
\item \textbf{PairRNN$_{\scriptscriptstyle \rm{LSTM}}$}: A variation of \projtitleLSTM, which directly feed node feature into a Pseudo-Siamese recurrent neural network layer with two LSTMs.
\item \textbf{G-PairRNN$_{\scriptscriptstyle \rm{LSTM}}$}: A variation of \projtitleLSTM, which neglects the heterogeneity of academic network and generates representation on the original graph.
\item \textbf{M-PairRNN$_{\scriptscriptstyle \rm{LSTM}}$}: A variation of \projtitleLSTM, which removes semantic-level attention layer and assigns the same importance to each meta-path.
\item \textbf{\projtitleLSTM}: 
The proposed model that fuses attribute, structure and semantic feature for node embedding generation with an semantic attention mechanism. 
\end{itemize}
\subsection{Implementation Details}
In Aminer-AND, the selected meta-paths of our method consist of \textit{Paper-Author-Paper}, \textit{Paper-Topic-Paper} and \textit{Paper-Venue-Paper}. 
We use the author's affiliation as the discriminative attribute to separate papers into small blocks and we use the same trainset and testset as in~\cite{zhang2018name}. 

In Semantic Scholar, the selected meta-paths of our method consist of \textit{Paper-Paper}, \textit{Paper-Author-Paper}, \textit{Paper-Topic-Paper}, and \textit{Paper-Venue-Paper}. 
We use the author's email as the discriminative attribute to separate papers into small blocks.
To evaluate the learning ability of models, we test them on Semantic Scholar with different training ratios $\{20\%, 40\%, 60\%, 80\%\}$. 

The common training parameters are set as learning rate = $5e-4$ and dropout = 0.2. 
% For all models, we randomly initialize parameters and optimize them with Adam\cite{kingma2014adam}. 
The node embedding dimension is set to 64 and the classifiers of all methods is a three-layer fully-connected neural network with a ReLU function.
In our proposed model \projtitleLSTM, $K$ is set to 2 and the dimension of meta-path preference vector $\mathbf{a}$ is set to 32. 
\subsection{Results and Discussions}
% The results on Aminer-AND and Semantic Scholar are reported in Table~\ref{table:result_Aminer} and Table~\ref{table:result_Semantic}, respectively. 
The performance of different methods on some sampled names of Aminer-AND is reported in Table~\ref{table:result_Aminer}. 
The results on Semantic Scholar is reported in Table~\ref{table:result_Semantic}.
Major findings from experimental results can be summarized as follows:
\begin{table*}[t]
\caption{The detailed results (\%) on Aminer-AND}
\label{table:result_Aminer}
\centering
\scriptsize
\begin{tabular}{l|m{0.61cm}<{\centering}m{0.61cm}<{\centering}m{0.61cm}<{\centering}|m{0.61cm}<{\centering}m{0.61cm}<{\centering}m{0.61cm}<{\centering}|m{0.61cm}<{\centering}m{0.61cm}<{\centering}m{0.61cm}<{\centering}|m{0.61cm}<{\centering}m{0.61cm}<{\centering}m{0.61cm}<{\centering}|m{0.61cm}<{\centering}m{0.61cm}<{\centering}m{0.61cm}<{\centering}}
\hline
                                & \multicolumn{3}{c|}{\textbf{Attr.}}     & \multicolumn{3}{c|}{\textbf{Struc.}}    & \multicolumn{6}{c|}{\textbf{Attr. + Struc.}}                          & \multicolumn{3}{c}{\textbf{Attr. + Struc. + Sem.}} \\ \cline{2-16} 
                                & \multicolumn{3}{c|}{\textbf{Louppe et al.}} & \multicolumn{3}{c|}{\textbf{Zhang et al.}} & \multicolumn{3}{c|}{\textbf{GHOST}} & \multicolumn{3}{c|}{\textbf{Aminer}} & \multicolumn{3}{c}{\textbf{\projtitleLSTM}}                   \\ \cline{2-16} 
\multirow{-3}{*}{\textbf{Name}} & Prec          & Rec          & F1           & Prec         & Rec          & F1           & Prec       & Rec        & F1        & Prec        & Rec        & F1        & Prec               & Rec               & F1                \\ \hline
\rowcolor{Gray} 
Hongbin Li                      & 19.48         & 85.96        & 31.77        & 54.66        & 53.05        & 53.84        & 56.29      & 29.12      & 38.39     & 77.20       & 69.21      & 72.99     & 88.89              & 65.98             & 75.74             \\
Hua Bai                         & 36.39         & 41.33        & 38.70        & 58.58        & 35.90        & 44.52        & 83.06      & 29.54      & 43.58     & 71.49       & 39.73      & 51.08     & 89.22              & 70.54             & 78.79             \\
\rowcolor{Gray}
Kexin Xu                        & 91.26         & 98.35        & 94.67        & 90.02        & 82.47        & 86.08        & 92.90      & 28.52      & 43.64     & 91.37       & 98.64      & 94.87     & 85.19              & 71.88             & 77.97             \\
Lu Han                          & 30.25         & 46.65        & 36.70        & 47.88        & 20.62        & 28.82        & 69.72      & 17.39      & 27.84     & 51.78       & 28.05      & 36.39     & 92.43              & 69.62             & 79.42             \\
\rowcolor{Gray}
Lin Huang                       & 24.86         & 71.32        & 36.87        & 71.84        & 34.17        & 46.31        & 86.15      & 17.25      & 28.74     & 77.10       & 32.87      & 46.09     & 88.26              & 73.44             & 80.17             \\
Meiling Chen                    & 58.32         & 47.14        & 52.14        & 59.36        & 28.80         & 38.79        & 86.11      & 23.85      & 37.35     & 74.93       & 44.70       & 55.99     & -                  & -                 & -                 \\
\rowcolor{Gray}
Min Zheng                       & 25.86         & 32.67        & 28.87        & 54.76        & 19.70        & 28.98        & 80.50      & 15.21      & 25.58     & 57.65       & 22.35      & 32.21     & 86.07              & 82.03             & 84.00                \\
Qiang Shi                       & 35.31         & 47.18        & 40.39        & 43.84        & 36.94        & 40.10        & 53.72      & 26.80      & 35.76     & 52.20       & 36.15      & 42.72     & 80.25              & 69.15             & 74.29             \\
\rowcolor{Gray}
Rong Yu                         & 38.85         & 91.43        & 54.53        & 65.48        & 40.85        & 50.32        & 92.00      & 36.41      & 52.17     & 89.13       & 46.51      & 61.12     & 90.67              & 68.69             & 78.16             \\
Tao Deng                        & 40.46         & 51.38        & 45.27        & 53.04        & 29.89        & 38.23        & 73.33      & 24.50      & 36.73     & 81.63       & 43.62      & 56.86     & 88.42              & 65.12             & 75.00                \\
\rowcolor{Gray}
Wei Quan                        & 37.86         & 63.41        & 47.41        & 64.45        & 47.66        & 54.77        & 86.42      & 27.80      & 42.07     & 53.88       & 39.02      & 45.26     & 75.76              & 78.13             & 76.92             \\
Xudong Zhang                    & 72.38         & 79.83        & 75.92        & 70.20        & 23.35        & 35.04        & 85.75      & 7.23       & 13.34     & 62.40        & 22.54      & 33.12     & -                  & -                 & -                 \\
\rowcolor{Gray}
Xu Xu                           & 22.55         & 64.40        & 33.40        & 48.16        & 41.87        & 44.80        & 61.34      & 21.79      & 32.15     & 74.18       & 45.86      & 56.68     & 78.68              & 79.08             & 78.88             \\
Yanqing Wang                    & 29.64         & 79.08        & 43.11        & 60.40        & 51.97        & 55.87        & 80.79      & 40.39      & 53.86     & 71.52       & 75.33      & 73.37     & 77.42              & 64.86             & 70.59             \\
\rowcolor{Gray}
Yong Tian                       & 32.08         & 63.71        & 42.67        & 70.74        & 56.85        & 63.04        & 86.94      & 54.58      & 67.06     & 76.32       & 51.95      & 61.82     & 87.80               & 70.59             & 78.26             \\ \hline
Average                          & 57.09         & 77.22        & 63.10        & 70.63        & 59.53        & 62.81        & 81.62      & 40.43      & 50.23     & 77.96       & 63.03      & 67.79     & \textbf{87.93}              & \textbf{77.74}             & \textbf{82.53}             \\ \hline
\end{tabular}
\end{table*}
\begin{table*}[t]
\caption{Quantitative results and standard deviation  (\%) on SemanticScholar}
\label{table:result_Semantic}
\centering
\scriptsize
\begin{tabular}{m{0.83cm}|m{0.6cm}|
m{1.1cm}<{\centering} |
m{1.5cm}<{\centering} |m{1.1cm}<{\centering}|m{1.1cm}<{\centering}|m{1.1cm}<{\centering}|m{1.1cm}<{\centering}|m{1.5cm}<{\centering}|
m{1.5cm}<{\centering}|
m{1.5cm}<{\centering}}
\arrayrulecolor{tabcolor}
\hline
                                    & \multicolumn{1}{c|}{}                                    & \multicolumn{2}{c|}{\textbf{Attr.}} & \multicolumn{5}{c|}{\textbf{Attr. + Struc.}}                                                                                      & \multicolumn{2}{c}{\textbf{Attr. + Struc. + Sem.}} \\ \arrayrulecolor{tabcolor}\cline{3-11} 
\multirow{-2}{*}{\textbf{Metrics}}  & \multicolumn{1}{c|}{\multirow{-2}{*}{\textbf{Training}}} & \textbf{MLP}                     & \textbf{PairRNN$_{\scriptscriptstyle \rm{LSTM}}$}                    & \textbf{Deepwalk} & \textbf{GraphSage} & \multicolumn{1}{c|}{\textbf{Aminer}} & \multicolumn{1}{c|}{\textbf{Kim et al.}} & \textbf{G-PairRNN$_{\scriptscriptstyle \rm{LSTM}}$} & \textbf{M-PairRNN$_{\scriptscriptstyle \rm{LSTM}}$}                         & \textbf{\projtitleLSTM}                         \\\hline
                                    & \textbf{20\%}                                            & 92.24$\pm$1.36                       & 94.78$\pm$0.74                           & 92.26$\pm$0.62        & 95.56$\pm$0.35         & 96.73$\pm$0.35                           & 96.88$\pm$0.46                               & 95.93$\pm$0.57          & 96.40$\pm$0.54                                  & \textbf{96.95}$\pm$\textbf{1.36}                          \\
                                    & \textbf{40\%}                                            & 93.88$\pm$1.01                       & 96.46$\pm$1.12                           & 93.85$\pm$0.65        & 96.27$\pm$0.18         & 96.59$\pm$0.33                           & 96.80$\pm$0.16                               & 96.34$\pm$0.61          & 96.73$\pm$0.69                                  & \textbf{97.01}$\pm$\textbf{0.45}                          \\
                                    & \textbf{60\%}                                            & 94.43$\pm$0.69                       & 97.34$\pm$1.08                           & 94.47$\pm$0.46        & 97.49$\pm$0.32         & 97.48$\pm$0.24                           & 97.54$\pm$0.35                               & 97.19$\pm$0.71          & 97.56$\pm$0.26                                  & \textbf{97.91}$\pm$\textbf{0.18}                          \\
\multirow{-4}{*}{\textbf{Accuracy}} & \textbf{80\%}                                            & 94.24$\pm$1.42                       & 97.56$\pm$0.26                           & 94.50$\pm$0.74        & 97.85$\pm$0.29         & 97.75$\pm$0.23                           & 97.38$\pm$0.23                               & 97.88$\pm$0.84          & 97.81$\pm$0.38                                  & \textbf{98.50$\pm$0.41}                          \\ \arrayrulecolor{tabcolor}\hline
                                    & \textbf{20\%}                                            & 92.14$\pm$1.49                       & 95.05$\pm$0.66                           & 92.37$\pm$0.54        & 95.62$\pm$0.35         & 96.84$\pm$0.32                           & 96.94$\pm$0.54                               & 96.10$\pm$0.53          & 96.54$\pm$0.50                                  & \textbf{97.04}$\pm$\textbf{1.30 }                         \\
                                    & \textbf{40\%}                                            & 93.91$\pm$1.00                       & 96.58$\pm$1.06                           & 93.92$\pm$0.59        & 96.33$\pm$0.17         & 96.66$\pm$0.33                           & 96.84$\pm$0.16                               & 96.48$\pm$0.57          & 96.84$\pm$0.64                                  & \textbf{97.12}$\pm$\textbf{0.43}                          \\
                                    & \textbf{60\%}                                            & 94.43$\pm$0.74                       & 97.40$\pm$1.05                           & 94.18$\pm$0.77        & 97.54$\pm$0.31         & 97.53$\pm$0.23                           & 97.59$\pm$0.34                               & 97.28$\pm$0.63          & 97.63$\pm$0.23                                  & \textbf{97.96}$\pm$\textbf{0.17}                          \\
\multirow{-4}{*}{\textbf{F1 Score}} & \textbf{80\%}                                            & 94.24$\pm$1.42                       & 97.66$\pm$0.27                           & 94.57$\pm$0.75        & 97.90$\pm$0.30         & 97.83$\pm$0.20                           & 97.42$\pm$0.24                               & 97.94$\pm$0.81          & 97.81$\pm$0.23                                  & \textbf{98.54$\pm$0.37}                          \\ \arrayrulecolor{tabcolor}\hline
                                    & \textbf{20\%}                                            & 92.24$\pm$1.36                       & 97.61$\pm$0.38                           & 92.26$\pm$0.62        & 96.10$\pm$1.93         & 98.02$\pm$0.59                           & 97.48$\pm$1.94                               & 92.85$\pm$9.56          & 97.96$\pm$1.13                                  & \textbf{98.12}$\pm$\textbf{1.08}                          \\
                                    & \textbf{40\%}                                            & 93.88$\pm$1.01                       & 95.38$\pm$5.23                           & 93.85$\pm$0.65        & 96.63$\pm$1.49         & 97.29$\pm$0.38                           & 95.39$\pm$5.72                               & 97.65$\pm$0.86          & 95.57$\pm$6.60                                  & \textbf{98.55}$\pm$\textbf{1.05}                          \\
                                    & \textbf{60\%}                                            & 94.43$\pm$0.69                       & 98.54$\pm$0.39                           & 94.27$\pm$0.85        & 97.91$\pm$0.86         & 98.32$\pm$0.56                           & 97.73$\pm$1.13                               & 98.86$\pm$0.43          & 99.07$\pm$0.30                                  & \textbf{99.31}$\pm$\textbf{0.45}                          \\
\multirow{-4}{*}{\textbf{AUC}}      & \textbf{80\%}                                            & 94.24$\pm$1.42                       & 98.43$\pm$0.57                           & 94.50$\pm$0.74        & 98.12$\pm$0.20         & 98.73$\pm$0.36                           & 97.70$\pm$0.59                               & 98.76$\pm$0.74          & 98.27$\pm$0.22                                  & \textbf{99.18}$\pm$\textbf{0.79}                          \\ \arrayrulecolor{tabcolor}\hline
\end{tabular}
\vspace{-0.6em}
\end{table*}

\textbf{Performance Comparison.} 
As shown in Table~\ref{table:result_Aminer} and Table~\ref{table:result_Semantic}, by incorporating attribute, structure and semantic information, \projtitleLSTM outperforms all baselines in both datasets. 
Generally, GNN based methods that combine the attribute and structure information usually perform better than those methods which only exploit attribute information. 
Compared to simply concatenate representations of nodes, the Pseudo-Siamese RNN network can better extract inherent relations of paper sequence. 
Compared to taking the graph as homogeneous, M-PairRNN$_{\scriptscriptstyle \rm{LSTM}}$ and \projtitleLSTM can exploit semantic information successfully and show their superiority. 
It demonstrates that combined use of attribute, structure, and semantic features better capture the similarities between papers. 
In addition, the semantic-level attention mechanism in \projtitleLSTM can exploit semantic information more properly. 
% In addition, incorporating meta-paths is essentially a better way to characterize the rich semantics and the semantic-level attention mechanism in \projtitleLSTM can exploit semantic information more properly. 

\begin{figure}[t]
% \vspace{-0.6em}
\centerline{\includegraphics[width=3.3in]{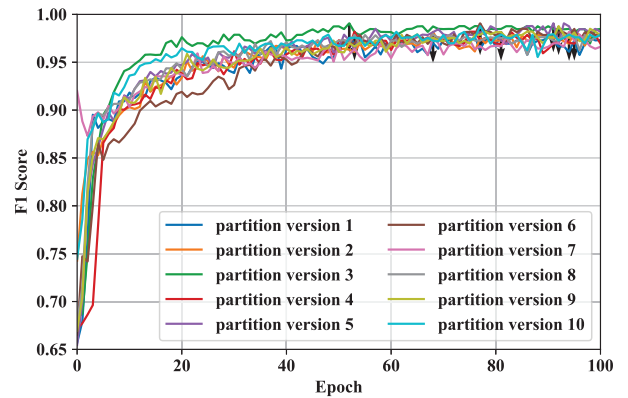}}
\caption{Performance of \projtitleLSTM on different Semantic Scholar partition version with training ratio of 80\%.}
\label{fig:epoch}
\vspace{-1em}
\end{figure}

Fig.~\ref{fig:epoch} shows F1 scores of \projtitleLSTM on different partition versions of Semantic Scholar with training ratio of 80\%. 
After adequate rounds of training, the performance of \projtitleLSTM on each dataset partition version has gained stability and certainty and is difficult to be further improved though fluctuations exist. 

\begin{figure}[t]
\centering 
\vspace{-0.6em}
\subfigure[F1 scores with different training ratio]{ 
\begin{minipage}[t]{1\linewidth} 
\centering
\includegraphics[width=3.1in]{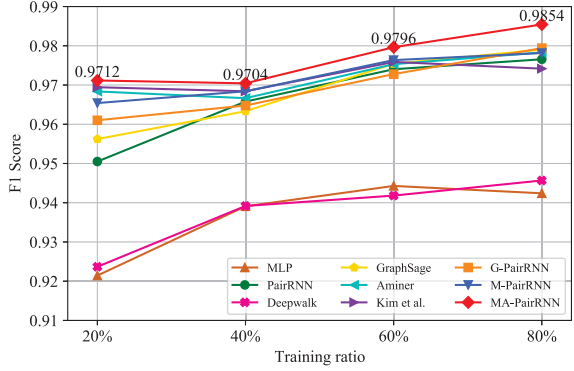}
\end{minipage} }%
\vspace{-0.5em}
\subfigure[Distributions of F1 scores with different training ratio]{ 
\begin{minipage}[t]{1\linewidth}
\centering
\includegraphics[width=3.3in]{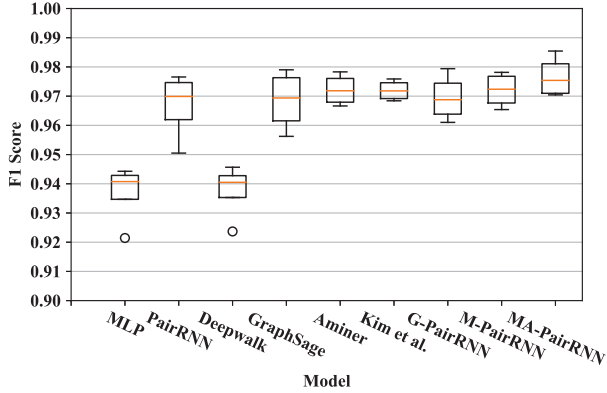} 
%\caption{fig2} 
\end{minipage} }%
\caption{Performance with different training ratio on Semantic Scholar.}
\vspace{-0.6em}
\label{fig:proportion}
\end{figure}

\textbf{Impact of training ratio.} 
F1 scores of all methods on Semantic Scholar with different training ratio are shown in Fig.~\ref{fig:proportion}~(a) and their distributions are shown in Fig.~\ref{fig:proportion}~(b). 
The performances of all methods get worse as the training ratio decrease. 
Our method \projtitleLSTM and its variants suffer less performance degradation than others, which shows better learning ability. 
% than other methods.

\textbf{Siamese Network v.s Pseudo-Siamese Network.}
As mentioned above, Pseudo-Siamese neural network component consists of two RNNs with different parameters. 
We also test three variations including a Pseudo-Siamese network with two BiLSTM  (MA-PairRNN$_{\scriptscriptstyle \rm{BiLSTM}}$), a Siamese network with two parameter-shared LSTM  (MA-RNN$_{\scriptscriptstyle \rm{LSTM}}$), and a Siamese network with two parameter-shared BiLSTM  (MA-RNN$_{\scriptscriptstyle \rm{BiLSTM}}$). 
Results on Semantic Scholar are shown in Table.~\ref{table:Siamese}.
We can see that Pseudo-Siamese Network models have a better performance than the other two Siamese Network models. 
Based on our assumption that papers during the period of discriminative attributes changing have similar text and structure features, the paper sequence published earlier is fed into RNN in publication time order and the other is in reverse order. 
Pseudo-Siamese network may better capture the changing trend of research topic and scholar relationship.
\begin{table}[t]
\caption{Performance comparison (\%) of different sequence representation model on Semantic Scholar}
\label{table:Siamese}
\centering
% \vspace{-0.6em}
\begin{tabular}{m{2.9cm}<{\centering}|m{1.3cm}<{\centering}|m{1.3cm}<{\centering}|m{1.3cm}<{\centering}}
\hline
\textbf{Models}        & \textbf{Accuracy}       & \textbf{F1 score}       & \textbf{AUC}            \\ \hline
\textbf{MA-PairRNN$_{\scriptscriptstyle \rm{LSTM}}$}   & \textbf{98.50} & \textbf{98.54} & \textbf{99.18} \\ \hline
\textbf{MA-PairRNN$_{\scriptscriptstyle \rm{BiLSTM}}$} & 98.47          & 98.52          & 99.17          \\ \hline
\textbf{MA-RNN$_{\scriptscriptstyle \rm{LSTM}}$}       & 97.88          & 97.96          & 99.00          \\ \hline
\textbf{MA-RNN$_{\scriptscriptstyle \rm{BiLSTM}}$}     & 98.25          & 98.28          & 99.17          \\ \hline
\end{tabular}
\vspace{-0.3em}
\end{table}

\begin{figure}[t]
\vspace{-0.6em}
\centerline{\includegraphics[width=3.3in]{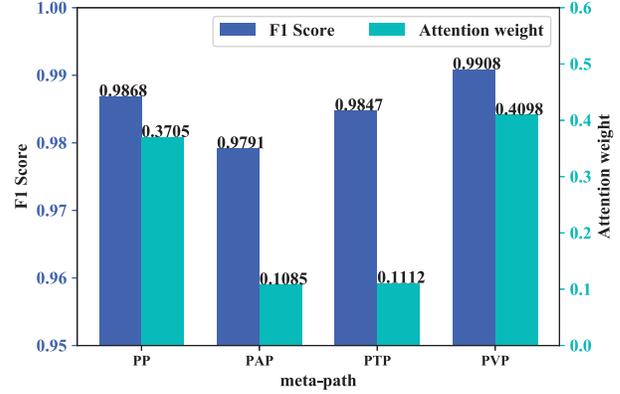}}
\caption{Performance of single meta-path and corresponding attention value.}
\vspace{-1em}
\label{fig:attention}
\end{figure}

\textbf{Impact of Different Meta-paths.} 
% As mentioned before, semantic-level attention can learn the importance of meta-paths. 
To verify the ability of semantic-level attention, we report F1 scores of \projtitleLSTM using single meta-path and corresponding attention values on Semantic Scholar in Fig.~\ref{fig:attention}.
Obviously, there is a positive correlation between the performance of each meta-path and its attention value. 
Among four meta-paths, \projtitleLSTM gives PVP the highest weight, which means that PVP is considered as the most critical meta-path in paper representation. 
It makes sense because author’s research areas are highly correlated with venues where their papers are published. 
% For example, computer vision researchers always publish papers in venues like \textit{CVPR} or \textit{ECCV}, whereas data mining researchers may submit their papers to \textit{KDD} or \textit{TKDE}. 
Meanwhile, PP is also given a high weight. 
It also makes sense because author's papers are often closely related and have similar references. 

\begin{figure}[t]
\vspace{-0.1em}
\centerline{\includegraphics[width=3.3in]{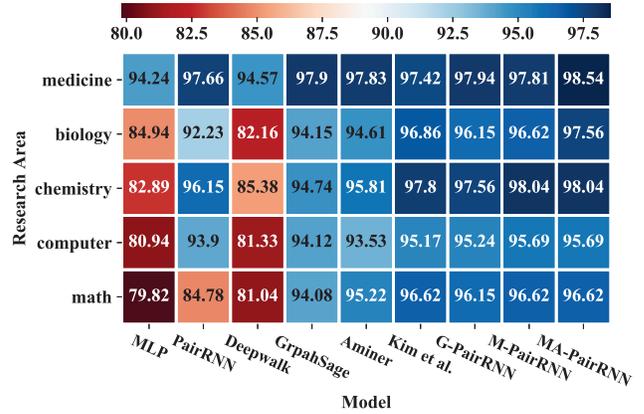}}
\centering
\caption{Performance (F1 score \%) in different research areas.}
\label{fig:area}
\vspace{-0.6em}
\end{figure}

\textbf{Generalization ability across research areas.}
On Semantic Scholar, our models are trained on papers of medical area. 
To verify the generalization ability of models across different research areas, we collected data of 100 authors from biology, chemistry, computer science, and mathematics area, respectively. 
The performance of all models on these data is shown in Fig.~\ref{fig:area}. 
When trained on data of the medical area and test on the other four areas, the performance degradations of our proposed model (\projtitleLSTM) and its variations (G-PairRNN$_{\scriptscriptstyle \rm{LSTM}}$ and M-PairRNN$_{\scriptscriptstyle \rm{LSTM}}$) are less than 3\%, which are better than other models. 
It indicates that the structure information can enhance model's generalization ability.
Most models perform better when transferred to biology and chemistry area than other two areas. 
It makes sense because these two areas share more area knowledge with the medical one.
\subsection{Parameters Analysis}
\begin{figure*}[t]
% \vspace{-0.6em}
\centering 
\subfigure[Dimension of the final node embedding $\mathbf{z}$]{ 
\begin{minipage}[t]{0.32\linewidth} 
\centering 
\includegraphics[width=2.3in]{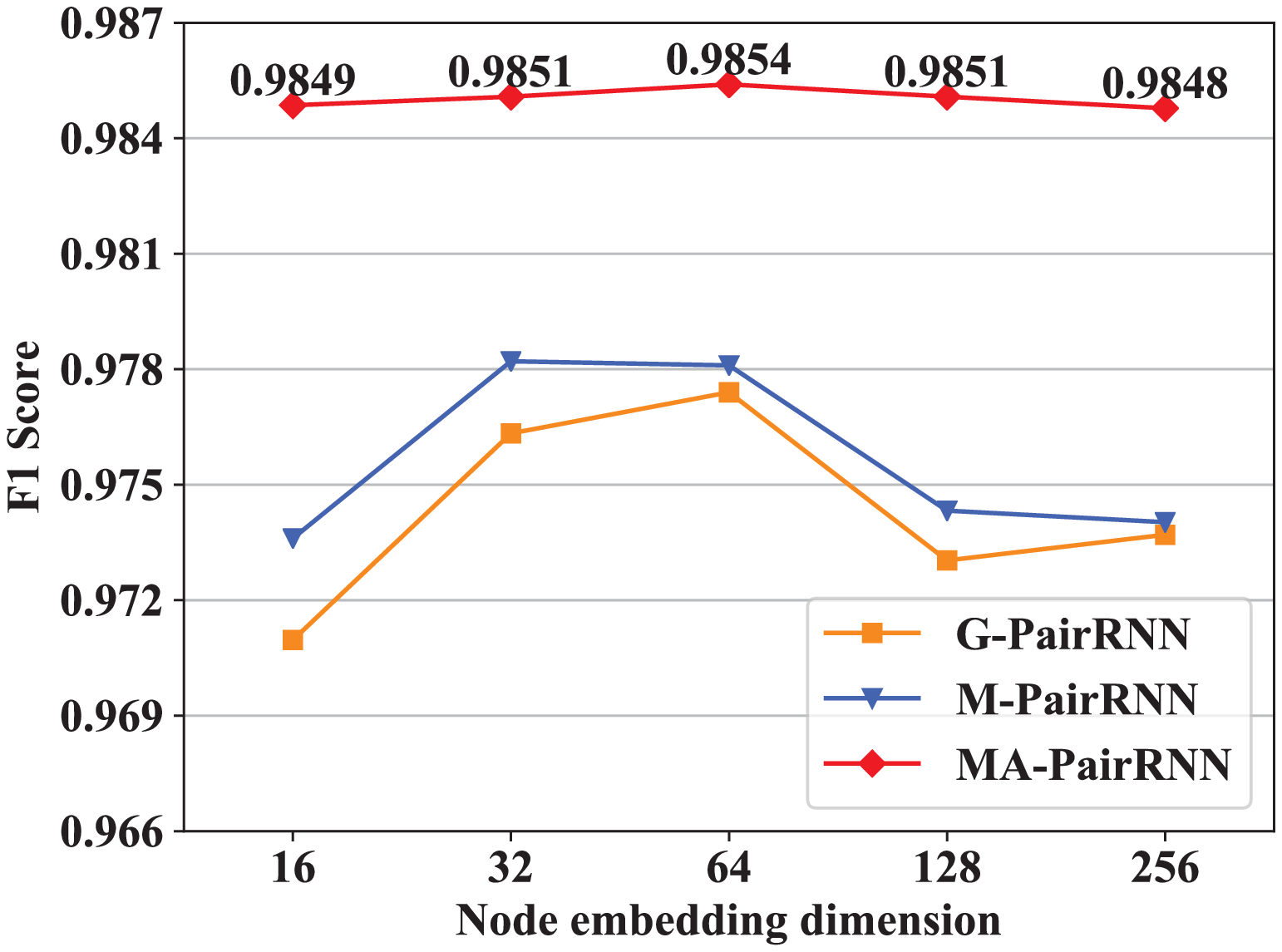} %
\end{minipage}% 
}% 
\subfigure[Dimension of semantic attention vector $\mathbf{a}$]{ 
\begin{minipage}[t]{0.32\linewidth} 
\centering \includegraphics[width=2.3in]{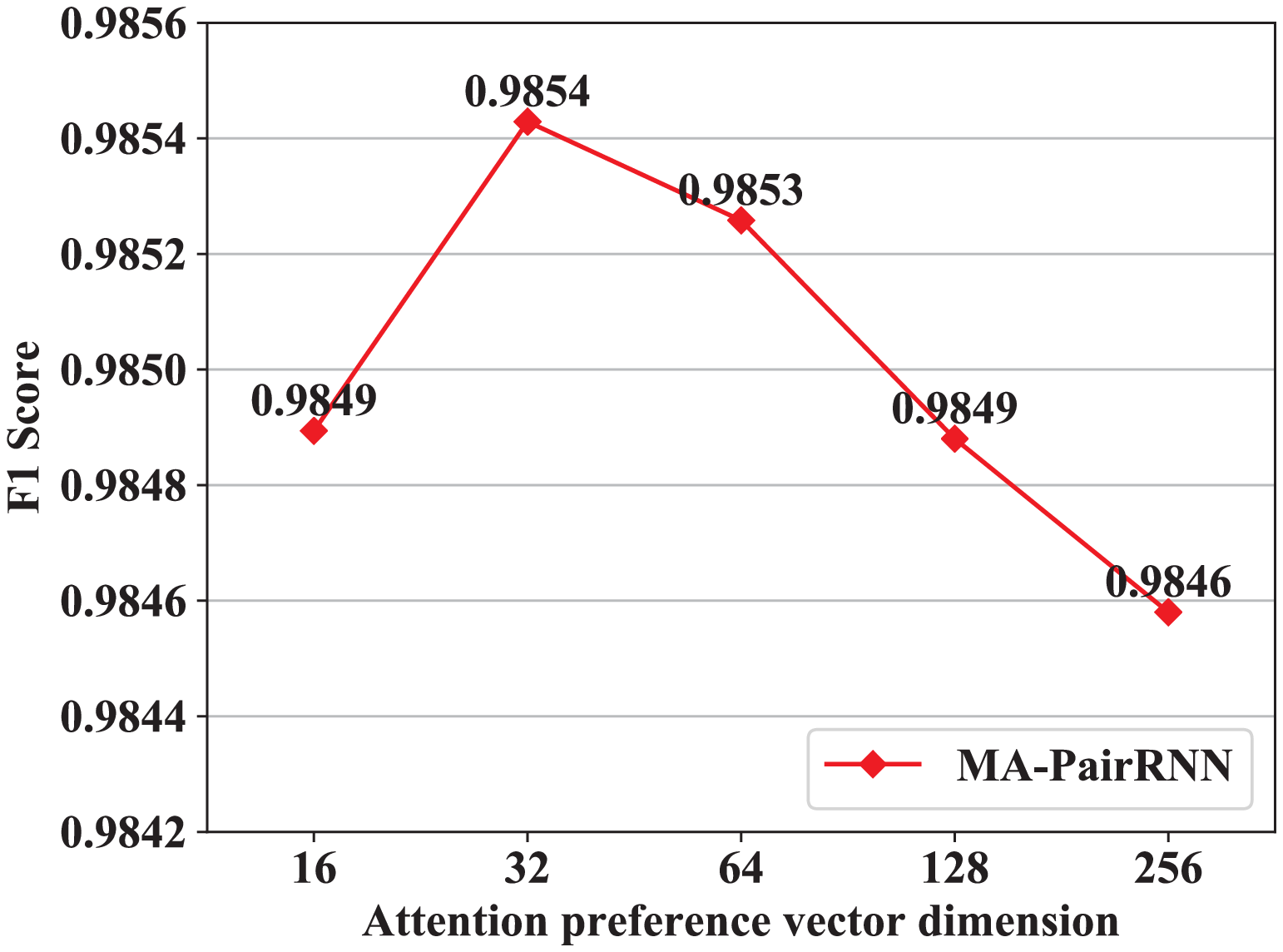} 
%\caption{fig2} 
\end{minipage} }% 
\subfigure[Coefficient $\eta$ of cosine similarity loss]{ 
\begin{minipage}[t]{0.32\linewidth} 
\centering 
\includegraphics[width=2.3in]{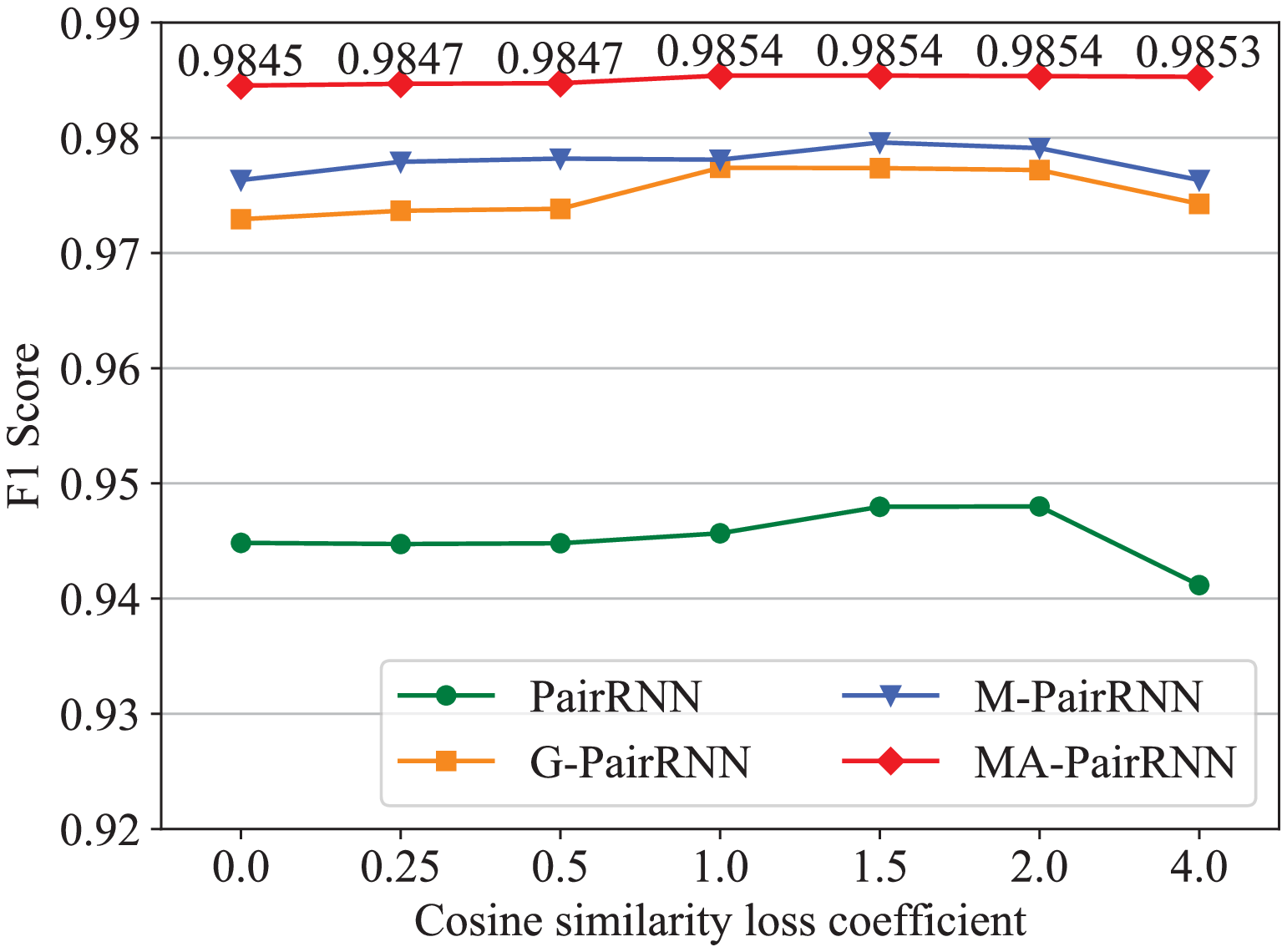} 
%\caption{fig2} 
\vspace{-0.6em}
\end{minipage} 
}% 
\centering 
\caption{Parameter sensitivity: Dimension of node embedding $\mathbf{z}$, Dimension of semantic attention vector $\mathbf{a}$ and Coefficient $\eta$ of cosine similarity loss.} 
\label{fig:sensitivity}
\end{figure*}
In this section, we will investigate how dimension of node embedding and attention preference vector and coefficient of similarity loss can affect classification performance. 
The results on Semantic Scholar are reported in Fig.~\ref{fig:sensitivity}. 
% Following previous experiment settings, we only change the analyzed parameter and report F1 scores on Semantic Scholar in Fig.~\ref{fig:sensitivity}. 

\textbf{Dimension of the final node embedding $\mathbf{z}$.}
The representation ability of graph embedding methods is affected by the dimension of node embedding $\mathbf{z}$. 
We explore its impact with various dimension \{16, 32, 64, 128, 256\}. 
As shown in Fig.~\ref{fig:sensitivity}~(a), the performance firstly improves with the increase of node embedding dimension, then degenerates slowly, and achieves the best performance at the dimension of 64. 
The reason may be that larger dimension could introduce some additional redundancies.

\textbf{Dimension of semantic attention vector $\mathbf{a}$.}
We evaluate the effect of semantic attention vector $\mathbf{a}$'s dimension in the set of $\{16, 32, 64, 128, 256\}$. 
% As shown in Fig.~\ref{fig:sensitivity}(b), the performance raises first and then starts to drop with the increase of the attention vector dimension. 
As shown in Fig.~\ref{fig:sensitivity}~(b), the F1 score has minor changes, which shows that \projtitleLSTM is not very sensitive to the dimension of attention preference vector.

\textbf{Coefficient $\eta$ of cosine similarity loss.}
% Besides above two parameters, our model also involves another important tuning parameter $\eta$ in \eqref{eq:loss}. 
The impact of similarity loss item is controlled by $\eta$. 
We vary $\eta \in \{0, 0.25, 0.5, 1, 1.5, 2, 4\}$. 
As shown in Fig.~\ref{fig:sensitivity}~(c), optimal performance is obtained near $\eta$ = 1, indicating that $\eta$ cannot be set too small or too large in order to prevent overfitting and underfitting. 
\subsection{Case Study}
We specifically choose three author variants named \textit{Jian Pei} in Semantic Scholar as a study case and we denote them as \textit{Jian Pei 1}, \textit{Jian Pei 2}, \textit{Jian Pei 3}. 
Statistics of selected three author variants are shown in Table.~\ref{table:case}. 
Our model classifies \textit{Jian Pei 1} and \textit{Jian Pei 2} as the same person while \textit{Jian Pei 3} is another person, which is consistent with the ground truth. 
We visualize the subgraph of the academic network that three author variants are in. 
% using Gephi\footnote{https://gephi.org/} in Fig.~\ref{fig:case}. 
The visualized subgraph includes papers and co-authors of the three author variants, and topics their papers related to. 
Papers of three author variants are colored blue, green, and red respectively and other nodes are colored by their type. 
Paper nodes of \textit{Jian Pei 1} colored blue and paper nodes of \textit{Jian Pei 2} colored green tend to be closely connected physically and many of them are connected by same topics (e.g., \textit{Data mining}, \textit{Social Network}) and same venues (e.g., \textit{KDD}, \textit{TKDE}). 
\textit{Jian Pei 3}'s paper nodes are connected to paper nodes of the other two by topic nodes such as \textit{Algorithm} and \textit{Simulation experiment}, which are used in many research areas.
\begin{table}[t]
\caption{Statics of selected author variants}
\label{table:case}
\centering
% \vspace{-0.7em}
\begin{tabular}{c|c|c|m{3.5cm}}
\hline
\textbf{author}     & \textbf{\#papers} & \textbf{\#citations} & \textbf{Most common topics} \\ \hline
                    &                   &                      & \textit{Data mining}                 \\
\textit{Jian Pei 1} & 441               & 23,729               & \textit{Social networks}             \\
                    &                   &                      & \textit{Frequent pattern mining}     \\ \hline
                    &                   &                      & \textit{Data mining}                 \\
\textit{Jian Pei 2} & 78                & 4,512                & \textit{Sequential pattern mining}   \\
                    &                   &                      & \textit{Frequent pattern mining}     \\ \hline
                    &                   &                      & \textit{Molecular synthesis}         \\
\textit{Jian Pei 3} & 36                & 690                  & \textit{Functional materials}        \\
                    &                   &                      & \textit{Convenient Syntheses}        \\ \hline
\end{tabular}
\vspace{-0.3em}
\end{table}
\begin{figure}[t]
\vspace{-0.3em}
\centerline{\includegraphics[width=2.9in]{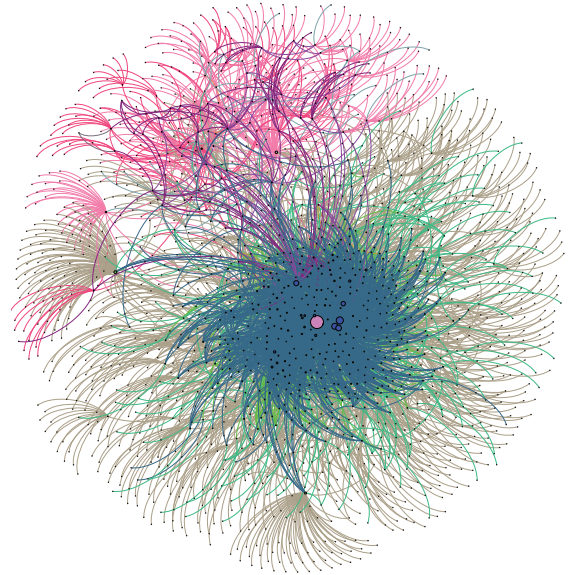}}
\caption{Subgraph visualization of selected author variants. Paper node color represents author variant  (Blue: \textit{Jian Pei 1}, Green: \textit{Jian Pei 2}, Red:  \textit{Jian Pei 3})}
\vspace{-0.6em}
\label{fig:case}
\end{figure}

%% file: 5_conclusion.tex
\vspace{-0.3em}
\section{CONCLUSION AND FUTURE WORK}
In this paper, we propose \projtitle, a novel pairwise node sequence classification framework for name disambiguation, in which multi-view graph embedding layer is designed to generate node representation inductively, and Pseudo-Siamese recurrent neural network is designed to learn sequence pair similarity. 
Our proposed method can learn node representation and sequence pair similarity simultaneously, and can scale to large graphs for its inductive capability. 
Experimental results on two real-world datasets demonstrate the effectiveness of our method. 
By analyzing the learned attention weights of meta-paths, \projtitle has proven its potentially good interpretability. 
By testing on data of unseen areas, \projtitle has also proven its good generalization ability. 
% There are some potential improvements to the proposed model that could be addressed in the future. 
% For example, the attention mechanism can also be adapted in sequence representation. 
In the future, we plan to leverage hierarchical clustering to address the problem that an author has diverse research areas and works with non-overlapping sets of co-authors corresponding to each research area.